# A single multi-configuration Direct Electron Detector for various electron imaging and diffraction-based techniques in SEM

**Nohayla El-Khairaoui[a], Julien Guyon[a], Nathalie Gey[a], Luc Morhain[a], Nabila Maloufi[a,*]**

[a] Université de Lorraine, CNRS, Arts et Métiers Institute of Technology, LEM3, F-57000 Metz, France.

*Corresponding author: nabila.maloufi@univ-lorraine.fr

## Graphical abstract

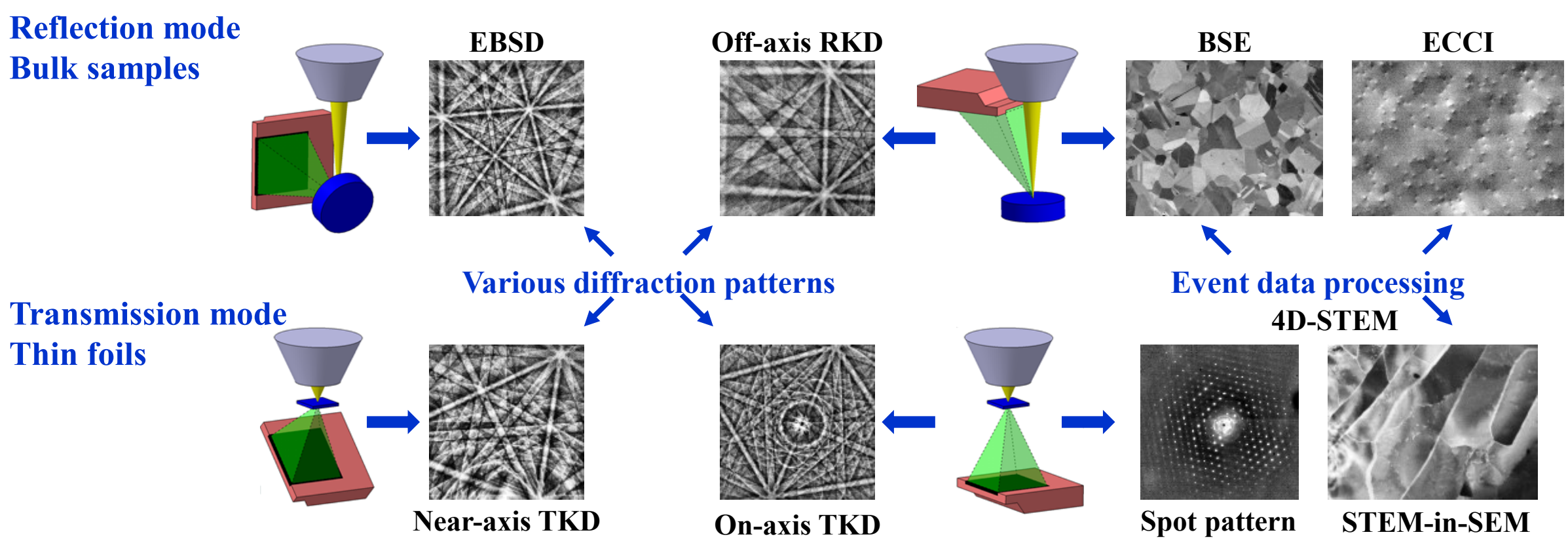

## Abstract

Addressing the need for efficient and integrated multiscale crystallographic and defect analyses of advanced materials, this paper presents the implementation of a new multi-configuration detection system, integrating a single Timepix3-based direct electron detector (DED) in a scanning electron microscope (SEM). By combining precise translation and rotation movements, this system enables, for the first time, the use of the same detector to realize all principal diffraction geometries. These include conventional Electron BackScatter Diffraction (EBSD), off-axis Reflexion Kikuchi Diffraction (RKD), and Transmission Kikuchi Diffraction (TKD) in on-, off- and near-axis configurations. Furthermore, transitions between all these geometries are accomplished without hardware modification. On the other hand, this work presents efficient reconstruction of electron images using the detector data-driven feature, extending thus its applicability to BackScattered Electron imaging (BSE), Electron Channelling Contrast Imaging (ECCI) and Scanning Transmission Electron Imaging in SEM (STEM-in-SEM) characterizations. High-quality Kikuchi patterns easily indexable were acquired across all geometries as well as micrographs of dislocations in both reflection and transmission modes. This is achieved thanks to the flexibility of the implemented detector, the optimizations made in acquisition parameters, such as energy filtering settings, and the efficiency of the developed custom approach used for electron data post-processing. Through this work, it is demonstrated that with a single DED assisted by an orientable support, it is possible to perform multiple advanced microstructural characterizations of both bulk samples and thin foils in the same SEM.

## Highlights:

- A multi-configuration SEM-based setup enabled by a single Timepix3 direct electron detector for advanced materials characterization.
- Novel off-axis RKD and near-axis TKD geometries using direct electron detector.
- First demonstration of BSE / ECCI using a direct electron detector.
- Saturation-free on-axis TKD via energy filtering and frame-based acquisition
- Efficient multiscale characterization of advanced materials including crystallographic and defect analysis.

# 1. Introduction

Scanning Electron Microscopy (SEM) is a cornerstone tool in materials science, enabling comprehensive microstructural characterization via various diffraction and imaging techniques. Among them Electron Backscatter Diffraction (EBSD) is the established standard for phase and crystal orientation characterization in bulk materials [1] while Electron Channelling Contrast Imaging (ECCI) constitute a powerful technique for lattice extended defects contrast and identification [2,3]. In recent years, Transmission Kikuchi Diffraction (TKD) extends SEM-based diffraction to thin samples, providing enhanced spatial resolution for nanostructured materials [4] while Scanning Transmission Electron Microscopy in a SEM (STEM-in-SEM) further expanded the analytic scope of SEM to nanostructures imaging [5]. Over time, TKD has evolved from off-axis to on-axis [6,7] and near-axis geometries to address challenges related to pattern distortion, angular coverage, and signal saturation. Conventionally, these techniques are implemented using dedicated detectors and specific geometrical configurations, each geometry is optimized for a single technique. While highly effective, this fragmented approach often requires hardware replacement or significant mechanical reconfiguration to access complementary information and thus introduces experimental complexity.

A major technological advance underpinning recent development in SEM diffraction has been the introduction of Direct Electron Detectors (DED) [8]. Unlike conventional scintillator-based cameras, DED detect electrons directly, offering enhanced signal-to-noise ratio (SNR) and superior dynamic range [9]. These capabilities enable improved detection of weak diffraction features and more flexible operation across beam energies and diffraction geometries. Among these detectors, the Timepix3 [10] hybrid pixel detector is particularly notable as it provides event-based acquisition through multiple modes, including Time-of-Arrival (ToA), thereby enabling time-resolved analysis of electron signals. This temporal information can be further exploited via advanced signal processing for complementary Backscattered Electron (BSE) imaging.

The advantages of using DED in SEM-based techniques was demonstrated by significant results obtained by several expert teams over the last years. Indeed, DED initial application in EBSD enabled the revelation of higher-order Kikuchi bands [11], while subsequent works demonstrated the benefits of electron energy filtering [12]. Furthermore, DED have facilitated the implementation of tilt-free EBSD, also known as reflection Kikuchi diffraction (RKD) [13–15], and were utilized for 3D EBSD applications [16]. In the context of TKD, DED have been used to achieve wider-angle [17] and high-resolution on-axis Kikuchi and spot patterns [18]. In this context the benefits of DED across these electron diffraction geometries were explored and compared [19]. Although the implementation of DED in SEM is booming as illustrated by these abundant works, many aspects remain to be investigated and enhanced as with any new development. While these detectors have become the leading technology for STEM-in-SEM analysis, particularly for 4D-STEM [20], their potential for near-axis TKD and ECCI has not yet been explored. Moreover, existing DED-based implementations continue to rely on multiple detectors or complex mechanical adjustments to transition between techniques and geometries, limiting experimental efficiency and continuity.

In this this work, we present a novel and highly versatile integration of a Timepix3-based detector within SEM. This unique implementation provides access to multiple diffraction and imaging techniques, using a single flexible detector without sensor replacement or major reconfigurations. In addition to EBSD, we demonstrate, in reflection mode and for the first time, the successful acquisition of indexable off-axis RKD patterns alongside high-contrast ECCI and BSE micrographs, all using a direct electron detector. The same detector also facilitates a full range of transmission modes, including the first reported use of a DED for near-axis TKD, as well as high-quality, non-saturated on-axis TKD patterns and STEM-in-SEM imaging. This multi-configuration setup enhances the functional versatility of SEM while reducing system complexity and operational cost, as a single detector can serve easily a wide range of techniques. This instrumental advancement, combined with the optimized data acquisition and post-treatment, establishes a multimodal and multi-scalable platform for efficient microstructural characterization.

## 2. Experimental methods

### 2.1 Concept and design of the multi-configuration detector

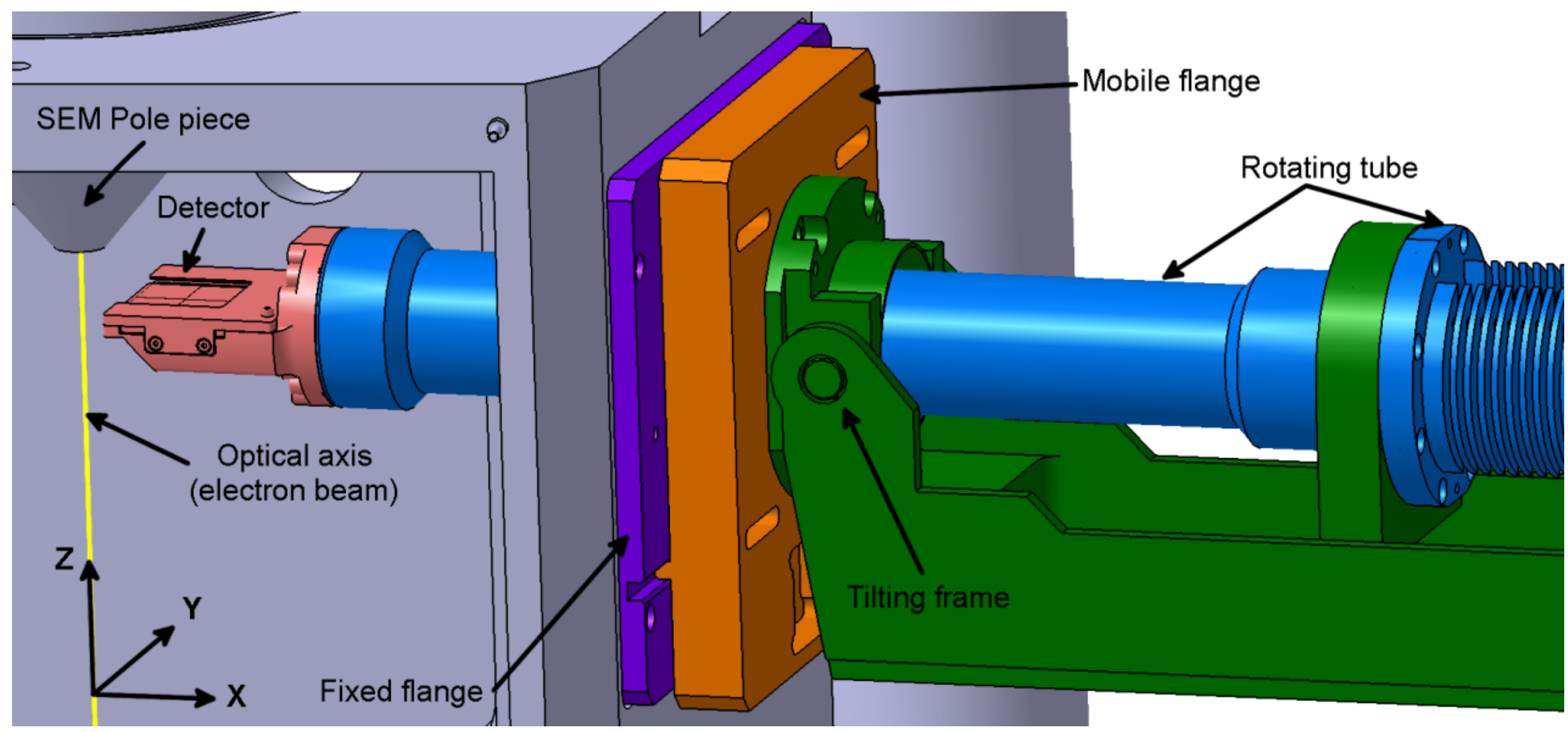

*Figure 1: CAD model of the detector integration within the SEM chamber along with the tube and the custom-built mobile flange. The sealing bellows is not shown in this representation for clarity.*

The custom-built detection system is centered around a Timepix3 direct electron detector (512×512 pixels, from Amsterdam Scientific Instruments). The sensor is mounted on a rotatable cylindrical tube attached to the SEM chamber port via a mobile flange. The detector assembly is integrated into a Zeiss Supra 40 SEM via a fixed flange (Figure 1). This design enables high flexibility in repositioning the detector in different geometries inside the SEM chamber thanks to its four degrees of freedom:

- Insertion (Tx): Standard insertion of the detector within the SEM chamber. It adjusts the proximity of the detector to the sample and pole piece along the X-axis.

- Tilt (Tz): Standard tilt mechanism around Y-axis, allowed by the tilting frame. It enables the displacement of the detector along Z-axis, and thus refining its Z position (height) within the chamber.

- Rotation (R): It is the novel integrated feature to the detector tube, allowing an easy rotation of the detector around X-axis. This enables the positioning of the detector in various orientations within the SEM chamber to comply with different characterization methods (EBSD, off-axis RKD, TKD).

- Lateral translation (Ty): This complementary feature is enabled by the custom mobile flange. It allows fine translational displacements along the Y-axis. These lateral movements allow the detector to be centred on the primary electron beam axis or shifted off-axis to prevent mechanical interference with the pole piece, and to vary the pole piece-detector distance.

## 2.2 Single detector: multiple diffraction and imaging applications

### 2.2.1 Diffraction configurations

The system flexibility arises from the integrated tube rotation (R) combined with Tx, Ty and Tz translations. This design allows the use of a single detector to accommodate various reflections and transmission geometries (Figure 2). Consequently, the setup enables the characterization of bulk samples via conventional EBSD (R= 90°) and off-axis RKD (R=0°). In this new RKD configuration, the detector is positioned near the pole piece (Figure 2, b) for tilt-free analysis of flat samples (tilt-free EBSD, BSE). The system also supports thin-foil characterization through off-axis (R=90°) and on-axis (R=180°) TKD. Notably, it introduces a novel near-axis TKD geometry (R=135°) by aligning the detector edge with the electron beam (Figure 2, e). All these diffraction configurations are performed seamlessly using the same detector within a single SEM.

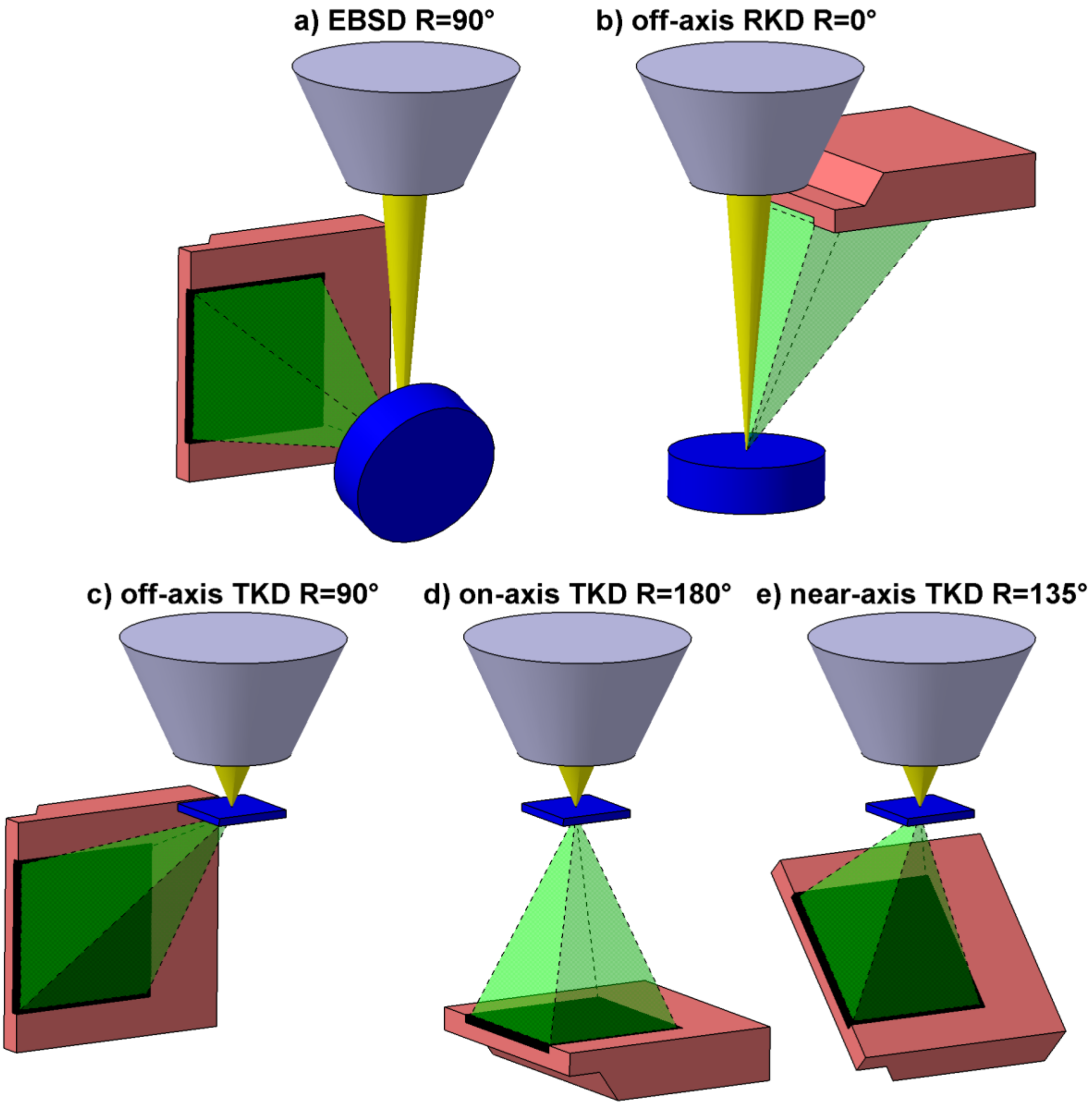

*Figure 2: Schematic representation of the five detector configurations. The setups for conventional EBSD, off-axis RKD, on-axis, off-axis, and near-axis TKD are shown, highlighting the relative positions of the pole piece (grey), incident electron beam (yellow), sample (blue), collected signal (green) and detector (brown) for each diffraction geometry.*

#### 2.2.2 BSE and STEM-in-SEM using time-resolved data

In addition to the acquisition of diffraction patterns using the frame-based mode, the Timepix3 detector enables the recording of individual events via the data-driven mode. In particular, X, Y detector coordinates and time-of-arrival of each detected electron are recorded. This acquisition approach paved the way for the reconstruction of electron images via in-house data post-processing algorithms. When configured for off-axis RKD, this detector provides simultaneous access to both electron diffraction patterns and BSE images. Moreover, the on-axis TKD geometry facilitates the concurrent acquisition of diffraction patterns and STEM images. This dual capability effectively enables 4D-STEM workflows within SEM. The proposed multi-configuration detector thus constitutes a SEM-based multimodal platform, integrating orientation microscopy and defect imaging in a single setup, thereby streamlining microstructural characterization.

## 3. Results and discussion

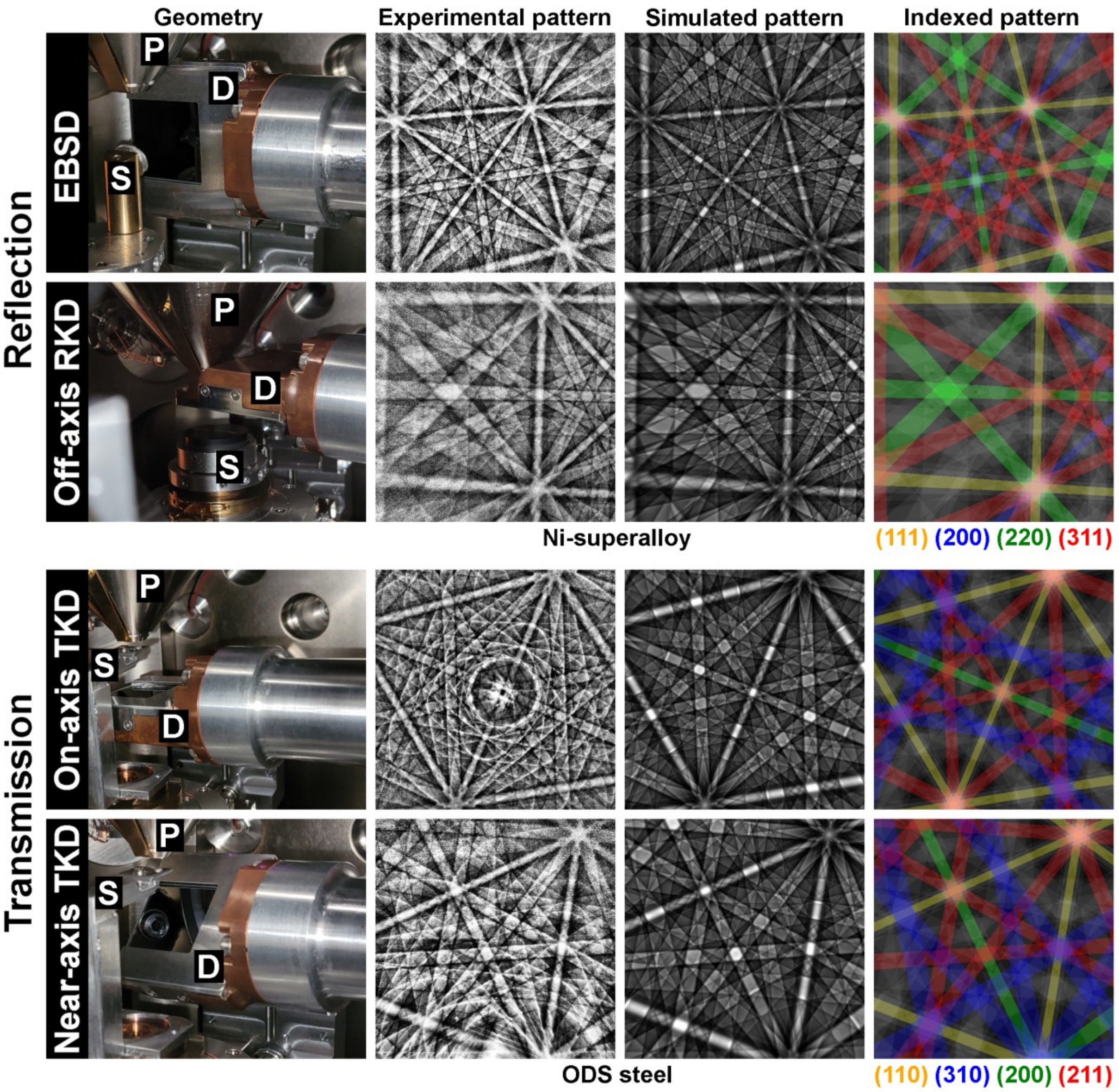

*Figure 3: Experimental setups (S: Sample, D: Detector, P: Pole piece), diffraction Kikuchi patterns acquired across the various geometries using the multi-configuration detector (Ni-superalloy bulk sample is used for EBSD and off-axis RKD; ODS steel thin foil used for on-axis and near-axis TKD), and the corresponding simulated and indexed patterns (via Bruker Esprit Dynamics software).*

The multi-configuration detector's performance was validated by acquiring high-quality diffraction patterns on bulk Ni-based superalloys in multiple reflection geometries as well as on a thin foil of Oxide Dispersion Strengthened (ODS) steel in multiple transmission geometries, using the counting frame-based mode (Figure 3). Optimized energy filtering, SEM and detector settings were applied to each geometry. Hence, reliable indexing and accurate crystal orientation determination was achieved for conventional EBSD and new challenging geometries such as off-axis RKD. Likewise, indexable near-axis TKD patterns were obtained using this flexible detector without the need for any complementary screen, as it is conventionally required. Remarkably, on-axis TKD patterns were acquired without central saturation of the electron beam. This superior quality results from a multi-frame acquisition approach using very short exposure time per frame combined with optimized acquisition parameters (low current and appropriate energy filtering). The short frames acquired are summed to enhance the total contrast and eliminate the central saturation, thus overcoming a major limitation encountered in conventional on-axis TKD.

Electron images in both reflection and transmission are reconstructed via advanced post-processing of raw event-based data (ToA). BSE micrographs of bulk Interstitial free steel (IF steel) reveals the polycrystalline microstructure of the material with clear contrast variations between grains (Figure 4, a), arising from differences in crystallographic orientation relative to the incident electron beam. It should be emphasised the high sensitivity to small grain internal misorientations, clearly resolving sub-boundaries and slight contrast variations within the grains (Figure 4, b). Moreover, ECCI micrographs clearly reveal in IF steel sample white fading traces corresponding to dislocations inclined relative to the sample surface, many of which are organized in to linear networks (Figure 4, d). Furthermore, dark/light spots indicative of threading dislocations are revealed in Gallium Nitride (GaN) sample (Figure 4, c). These micrographs were carried out at very short dwell time (DW=25.6 µs), achieving a spatial resolution comparable to that obtained with conventional dedicated detectors [21]. STEM-in-SEM micrograph (Figure 4, e) showcases the microstructure of the ODS steel thin foil, resolving small grains, sub-boundaries, and dislocations with high sensitivity to internal misorientations. These results demonstrate the effectiveness of the in-house reconstruction algorithm in leveraging the temporal resolution of the detector (1.52 ns) to resolve the subtle contrast modulations, induced by crystal misorientation and defects, in both backscattered and transmitted electron signals.

Using the same on-axis TKD geometry and acquisition parameters, and under the same SEM conditions, clear spot patterns were also obtained from a very thin region of the ODS steel thin foil (Figure 4, f). By recording the (X, Y) detector coordinates and ToA of every electron, the full phase space of the transmitted signal is captured which enables post-acquisition 4D-STEM analysis. Moreover, virtual bright-field and dark-field images can be reconstructed through selective spatial filtering of the raw event stream. This approach can effectively be extended to reflection-based analyses. By post-processing a single raw data file acquired in off-axis RKD, both the diffraction pattern (by processing the X-Y detector coordinates of each electron) and electron images (using ToA) can be reconstructed. Furthermore, by applying appropriate spatial selection of backscattered electrons from specific channelling regions of the pattern, high-contrast ECCI micrographs of dislocations can be generated entirely through post-processing. Overall, these results confirm that the implemented direct electron detector can enable both electron imaging and diffraction in both reflection and transmission modes, facilitating a comprehensive multiscale and multimodal characterization of advanced materials.

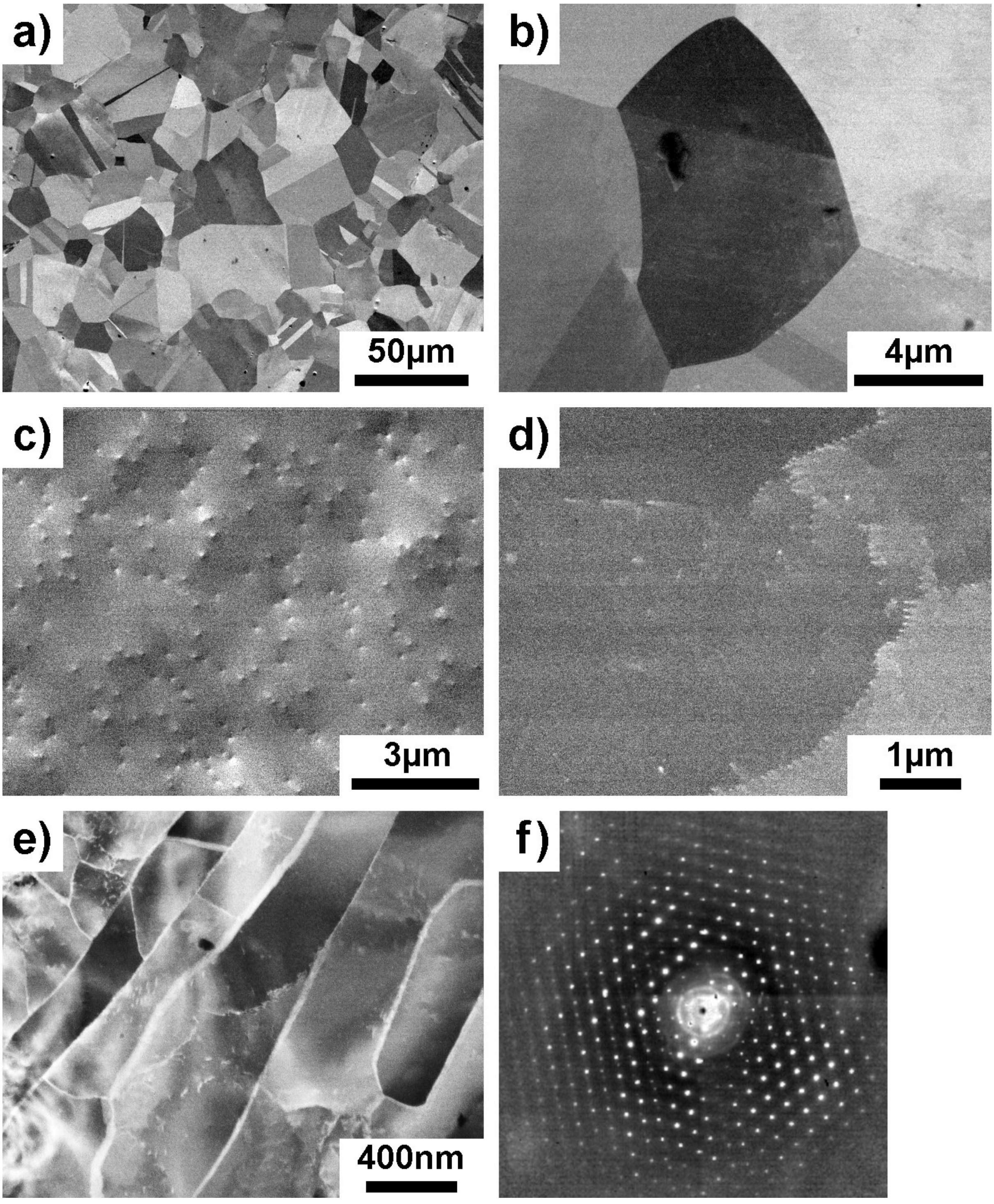

*Figure 4: Reconstructed images using Timepix3 electron data: (a) and (b) BSE micrographs of IF steel at different magnifications, (c) and (d) ECCI micrographs from GaN and IF steel bulk samples, respectively, both acquired at 20kV with a total scan time, e) STEM-in-SEM micrograph from ODS steel thin foil acquired at 30kV. (f) diffraction spots pattern acquired in frame-based mode from a very thin region of the same thin foil.*

## 4. Conclusion and perspectives

This publication demonstrates the successful implementation of a Timepix3-based direct electron detector within a unified multi-configuration SEM setup. The presented detector design enables seamless switching between multiple diffraction geometries and imaging modes without hardware modifications. Conventional configurations like EBSD and on-axis TKD were successfully performed. Moreover, the system enabled challenging new geometries, such as off-axis RKD and near-axis TKD, thereby expanding the range of SEM electron diffraction possibilities. Through appropriate energy filtering and a per-frame acquisition strategy, high-quality diffraction patterns were obtained across all configurations from diverse materials, notably overcoming saturation in the on-axis TKD geometry.

In addition to electron diffraction and STEM-in-SEM imaging, this work marks the first use of a direct electron detector for BSE and ECCI characterizations. The reconstructed micrographs reveal microstructural features including dislocation networks with high contrast, enabling the analysis of defect distributions in crystalline materials. The ability to acquire indexable diffraction patterns alongside high-contrast electron imaging micrographs, in both reflection and transmission, emphasizes the versatility of the detector setup. Overall, the results show that this single detector setup can effectively replaces a conventional ensemble of dedicated BSE, EBSD, TKD, and STEM-in-SEM detectors.

Beyond reducing system complexity, this multi-configuration detector broadens the scope of comprehensive microstructural analysis by enabling the seamless correlation of orientation mapping and electron imaging within a single configuration, using the same SEM and detector settings. This approach is particularly relevant for investigating microstructure-property relationships, where crystallographic information and defect structures must be analysed consistently from microscale to nanoscale. Future developments in real-time reconstruction, mapping and indexing will further enhance analytical throughput. This work positions the multi-configuration detector combined with the advancements in data acquisition and post-processing as a powerful solution for next-generation SEM characterizations in materials science.

### Acknowledgements

We acknowledge the experimental facilities MécaRhéo / MicroMat / Procédés from LEM3 (Université de Lorraine - CNRS UMR 7239).

**CRediT authorship contribution statement**

**Nohayla El-Khairaoui:** Conceptualization, Data curation, Formal analysis, Investigation, Methodology, Writing – original draft**. Julien Guyon:** Conceptualization, Formal analysis, Investigation, Methodology, Supervision, Validation, Writing – review & editing. **Nathalie Gey:** Funding acquisition, Investigation, Methodology, Resources, Supervision, Validation, Writing – review & editing. **Luc Morhain:** Conceptualization, Validation**. Nabila Maloufi**: Funding acquisition, Investigation, Methodology, Project administration, Resources, Supervision, Validation, Writing – review & editing.

## Declaration of competing interest

The authors declare no competing financial interests.

## Declaration of generative AI in scientific writing

During the preparation of this work the authors used AI tools for language editing and grammar correction only. After using these tools, the authors reviewed and edited the content as needed and take full responsibility for the content of the published article.

## References

[1] A.J. Schwartz, M. Kumar, B.L. Adams, D.P. Field, eds., Electron Backscatter Diffraction in Materials Science, Springer US, Boston, MA, 2009. https://doi.org/10.1007/978-0-387-88136-2.

[2] M.A. Crimp, Scanning electron microscopy imaging of dislocations in bulk materials, using electron channeling contrast, Microscopy Res & Technique 69 (2006) 374–381. https://doi.org/10.1002/jemt.20293.

[3] H. Mansour, M.A. Crimp, N. Gey, X. Iltis, N. Maloufi, Dislocation analysis of a complex subgrain boundary in UO2 ceramic using accurate electron channelling contrast imaging in a scanning electron microscope, Ceramics International 45 (2019) 18666–18671. https://doi.org/10.1016/j.ceramint.2019.06.091.

[4] R.R. Keller, R.H. Geiss, Transmission EBSD from 10 nm domains in a scanning electron microscope, Journal of Microscopy 245 (2012) 245–251. https://doi.org/10.1111/j.1365-2818.2011.03566.x.

[5] STEM-in-SEM: A re-emerging material measurement approach - 2022 - Wiley Analytical Science, Analytical Science Article DO Series (n.d.). https://analyticalscience.wiley.com/content/article-do/stem-in-sem-re-emerging-material-measurement-approach (accessed March 19, 2026).

[6] F. Niessen, A. Burrows, A.B.D.S. Fanta, A systematic comparison of on-axis and off-axis transmission Kikuchi diffraction, Ultramicroscopy 186 (2018) 158–170. https://doi.org/10.1016/j.ultramic.2017.12.017.

[7] E. Brodu, E. Bouzy, J.-J. Fundenberger, Diffraction contrast dependence on sample thickness and incident energy in on-axis Transmission Kikuchi Diffraction in SEM, Ultramicroscopy 181 (2017) 123–133. https://doi.org/10.1016/j.ultramic.2017.04.017.

[8] G. McMullan, A.R. Faruqi, R. Henderson, Direct Electron Detectors, in: Methods in Enzymology, Elsevier, 2016: pp. 1–17. https://doi.org/10.1016/bs.mie.2016.05.056.

[9] B.D.A. Levin, Direct detectors and their applications in electron microscopy for materials science, J. Phys. Mater. 4 (2021) 042005. https://doi.org/10.1088/2515-7639/ac0ff9.

[10] T. Poikela, J. Plosila, T. Westerlund, M. Campbell, M.D. Gaspari, X. Llopart, V. Gromov, R. Kluit, M.V. Beuzekom, F. Zappon, V. Zivkovic, C. Brezina, K. Desch, Y. Fu, A. Kruth, Timepix3: a 65K channel hybrid pixel readout chip with simultaneous ToA/ToT and sparse readout, J. Inst. 9 (2014) C05013–C05013. https://doi.org/10.1088/1748-0221/9/05/C05013.

[11] A.J. Wilkinson, G. Moldovan, T.B. Britton, A. Bewick, R. Clough, A.I. Kirkland, Direct Detection of Electron Backscatter Diffraction Patterns, Phys. Rev. Lett. 111 (2013) 065506. https://doi.org/10.1103/PhysRevLett.111.065506.

[12] S. Vespucci, A. Winkelmann, G. Naresh-Kumar, K.P. Mingard, D. Maneuski, P.R. Edwards, A.P. Day, V. O'Shea, C. Trager-Cowan, Digital direct electron imaging of energy-filtered electron backscatter diffraction patterns, Phys. Rev. B 92 (2015) 205301. https://doi.org/10.1103/PhysRevB.92.205301.

[13] T. Vystavěl, P. Stejskal, M. Unčovský, C. Stephens, Tilt-free EBSD, Microsc Microanal 24 (2018) 1126–1127. https://doi.org/10.1017/S1431927618006116.

[14] A.L. Marshall, J. Holzer, P. Stejskal, C.J. Stephens, T. Vystavěl, M.J. Whiting, The EBSD spatial resolution of a Timepix-based detector in a tilt-free geometry, Ultramicroscopy 226 (2021) 113294. https://doi.org/10.1016/j.ultramic.2021.113294.

[15] J. Holzer, A. Marshall, P. Stejskal, C. Stephens, T. Vystavěl, Large area EBSD mapping using a tilt-free configuration and direct electron detection sensor, Microsc Microanal 27 (2021) 1832–1835. https://doi.org/10.1017/S143192762100670X.

[16] K.P. Mingard, M. Stewart, M.G. Gee, S. Vespucci, C. Trager-Cowan, Practical application of direct electron detectors to EBSD mapping in 2D and 3D, Ultramicroscopy 184 (2018) 242–251. https://doi.org/10.1016/j.ultramic.2017.09.008.

[17] T. Zhang, T. Britton, Multi-exposure diffraction pattern fusion applied to enable wider-angle transmission Kikuchi diffraction with direct electron detectors, Ultramicroscopy 257 (2024) 113902. https://doi.org/10.1016/j.ultramic.2023.113902.

[18] S. Vespucci, A. Winkelmann, K. Mingard, D. Maneuski, V. O'Shea, C. Trager-Cowan, Exploring transmission Kikuchi diffraction using a Timepix detector, J. Inst. 12 (2017) C02075–C02075. https://doi.org/10.1088/1748-0221/12/02/C02075.

[19] T. Zhang, L. Berners, J. Holzer, T.B. Britton, Comparison of Kikuchi diffraction geometries in the scanning electron microscope, Materials Characterization 222 (2025) 114853. https://doi.org/10.1016/j.matchar.2025.114853.

[20] D. Chatterjee, J. Wei, A. Kvit, B. Bammes, B. Levin, R. Bilhorn, P. Voyles, An Ultrafast Direct Electron Camera for 4D STEM, Microsc Microanal 27 (2021) 1004–1006. https://doi.org/10.1017/S1431927621003809.

[21] H. Mansour, J. Guyon, M.A. Crimp, N. Gey, B. Beausir, N. Maloufi, Accurate electron channeling contrast analysis of dislocations in fine grained bulk materials, Scripta Materialia 84–85 (2014) 11–14. https://doi.org/10.1016/j.scriptamat.2014.03.001.